\documentclass[preprint,aps,prl,showpacs]{revtex4}

\usepackage[dvips]{graphicx}
\usepackage{amsmath}
\usepackage{SIunits}

\begin{document}

\title[Entanglement of two atoms using Rydberg blockade]{Analysis of the entanglement between two individual atoms using global Raman rotations}
\author{A. Ga\"{e}tan, C. Evellin, J. Wolters, P. Grangier, T. Wilk and A. Browaeys}
\address{Laboratoire Charles Fabry, Institut d'Optique, CNRS, Univ Paris-Sud,
Campus Polytechnique, RD 128,
91127 Palaiseau cedex, France}

\date{\today}

\begin{abstract}
Making use of the Rydberg blockade, we generate entanglement between two atoms
individually trapped in two optical tweezers. In this paper we detail the analysis of the
data and show that we can determine the amount of entanglement between the atoms
 in the presence of atom losses during the entangling sequence. Our model takes into account states outside the qubit basis and allows us to perform a partial reconstruction of the density matrix describing the two atom state. With this method we extract the  amount of entanglement between pairs of atoms still trapped after the entangling sequence and measure the fidelity with respect to the expected Bell state. We find a fidelity $F_{\rm pairs} =0.74(7)$ for the 62\% of atom pairs remaining in the traps at the end of the entangling sequence.
\end{abstract}


\maketitle

\section{Introduction}

Entanglement between two particles can be generated by designing and manipulating
interactions between them. For example, the entanglement in ion systems relies on the Coulomb interaction between the ions~\cite{BlattWinelandNat08}.
Entanglement is therefore difficult to produce in neutral atom systems, due to their weaker interactions. One solution, implemented in the first demonstration of  entanglement between 
neutral atoms, makes use of a high-Q cavity to mediate the interaction
between transient atoms~\cite{Hagley97}.
Another more recent approach uses ultra-cold atoms in optical lattices and the short-range s-wave interaction that occurs when their wavepackets overlap. This leads to the preparation of entangled states of a chain of atoms~\cite{Mandel03} or
pairs of atoms~\cite{Anderlini07}. This approach requires ground state cooling of atoms in their trapping potential and the ability to overlap their wavepackets during a controllable amount of time.
Furthermore, although there has been tremendous progress in this direction recently~\cite{Bakr09}, it is not easy to address atoms in optical lattices with a spacing between the wells of less than a micrometer.
An alternative approach is to store atoms in traps that are separated by
several micrometers in order to have addressability using standard optical techniques, and to avoid motional control of the atoms~\cite{Bergamini04,Nelson07}. One then needs an interaction which can act at long distance. Atoms in Rydberg states do provide such a long range interaction, which can reach several MHz at a distance of 10 micrometers. Moreover, this interaction can be switched on and off at will by placing the atoms in a Rydberg state for a controllable amount of time.
This approach using Rydberg interaction  has been proposed theoretically as a
way to perform fast quantum gates~\cite{Jaksch00, Lukin01, Saffman05} and
is intrinsically deterministic and scalable to more than two atoms.
Recent proposals extend this idea to the generation of various
entangled states~\cite{Moller08, Mueller09}.

Recently, two experiments implemented Rydberg interactions to demonstrate a cNOT
gate~\cite{Isenhower09} and to generate entanglement between two atoms trapped in optical tweezers~\cite{Wilk09}. In the present paper,  we analyze in detail the experiment of reference~\cite{Wilk09}. We explain how we extract the amount of entanglement with 
a method  based on global rotations of the state of the atoms.

The paper is organized as follows. In section~\ref{sectionblockade} we present the principle of the experiment. In section~\ref{sectionsetup} we detail the setup as well as the experimental sequence. We show that some  atoms are lost during the sequence. In section~\ref{sectiontheory} we present  the model  used to extract the amount of entanglement, which takes into account the loss of atoms.  In the following sections we present the experimental results:
in section~\ref{sectionlosses} we quantify  the atom losses and in section~\ref{sectionfidelity} we describe the partial tomography of the density matrix, extract the value of the fidelity and discuss the factors limiting this value.

\section{Rydberg blockade and entanglement}\label{sectionblockade}

The principle of the experiment relies on the Rydberg blockade effect demonstrated recently with two single atoms~\cite{Urban09, Gaetan09}. Due to their large electric dipole when they are in a Rydberg state $|r\rangle$, two atoms $a$ and $b$ interact strongly if they are close enough. This interaction leads to a shift $\Delta E$ of the doubly excited state $|r,r\rangle$.

As a consequence, a laser field coupling a ground state $|\!\uparrow\rangle$ and a Rydberg state $|r\rangle$ (with Rabi frequency $\Omega_{\uparrow r}$) cannot excite both atoms at the sameÊ time, provided that the linewidth of the excitation is smaller than $\Delta E$. In this blockade regime, the two-atom system behaves like an effective two-level system~\cite{Gaetan09}:
the ground state $|\!\uparrow,\uparrow\rangle$ is coupled to the excited state
\begin{equation}\label{eqno1}
|\Psi_{\rm r}\rangle = \frac{1} {\sqrt{2}}
(e^{i\mathbf{k}\cdot\mathbf{r}_a}|r,\uparrow\rangle +
e^{i\mathbf{k}\cdot\mathbf{r}_{b}}|\!\uparrow,r\rangle),
\end{equation}
where $\mathbf{k}=\mathbf{k}_{\rm R}+\mathbf{k}_{\rm B} $ is the sum of the wave vectors of the red (R) and blue (B) 
lasers used for the two-photon excitation (see section~\ref{sectionsetup} and figure~\ref{figure1}b)
and $\mathbf{r}_{a/b}$ are the positions of the atoms.
The coupling strength between these states is enhanced by a factor $\sqrt{2}$ with respect to the one between $|\!\uparrow\rangle$ and $|r\rangle$ for a single atom~\cite{Gaetan09}. Thus, starting from $|\!\uparrow,\uparrow\rangle$, a pulse of duration $\pi/(\sqrt{2}\,\Omega_{\uparrow r})$ prepares the state $|\Psi_{\rm r}\rangle$.
To produce entanglement between the atoms in two ground states, the Rydberg state
$|r\rangle$ is mapped onto another ground state $|\!\downarrow\rangle$ using the same blue laser
and  an additional red laser (wave vector $\mathbf{k'}_{\rm R}$)  
with a pulse of duration $\pi/\Omega_{r \downarrow}$ ($\Omega_{r \downarrow}$ is the two-photon Rabi frequency). This sequence results in the entangled state
\begin{equation}\label{eqno2}
|\Psi\rangle = \frac{1} {\sqrt{2}} (|\!\downarrow,\uparrow\rangle + e^{i \phi}|\!\uparrow,\downarrow\rangle),
\end{equation}
with $\phi = (\mathbf{k}_{\rm R} -\mathbf{k'}_{\rm R})\cdot (\mathbf{r}_{b} - \mathbf{r}_{a}) $,
assuming that the positions of the atoms are frozen~\footnote{For a
discussion of the case where atoms move, see reference~\cite{Wilk09}.}.
As the light fields are propagating in the same direction and the energy
difference between the two ground states is small, $\mathbf{k}_{\rm R}\simeq \mathbf{k'}_{\rm R}$.
This procedure therefore generates in a deterministic way  the well defined entangled state with $\phi =0$, which is the $|\Psi^+\rangle$ Bell state.

\section{Experimental setup and procedure}\label{sectionsetup}
\begin{figure}
\begin{center}
\includegraphics[width=9cm]{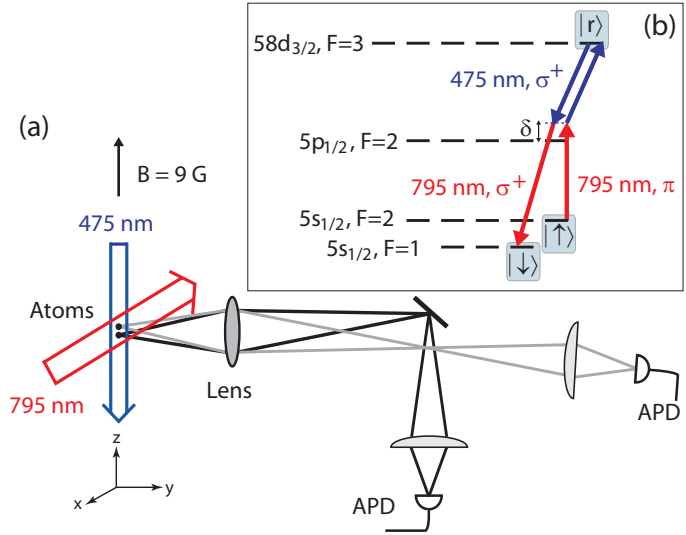}
\caption{
(a) Experimental setup. Two atoms are held at a distance of 4~$\mu$m in two optical tweezers formed by focused laser beams at 810~nm (not shown). The fluorescence of each atom is directed onto separate avalanche photodiodes (APDs). The $\sigma^+$-polarized 475~nm laser has a waist of 25~$\mu$m and is directed along the z-axis, the two 795~nm lasers have waists of 130~$\mu$m, copropagate along the x-axis and have both linear polarization, one along the quantization axis, the other perpendicular. The 475~nm and 795~nm lasers have powers of 30~mW and 15~mW, respectively, which correspond to Rabi frequencies $\Omega_B/(2\pi) \sim 25$~MHz and $\Omega_R/(2\pi) \sim 300$~MHz.
(b) Atomic level structure and lasers used for the excitation towards the Rydberg state. The 475~nm laser and the two 795~nm lasers are tuned to the two photon transitions from $|\!\uparrow\rangle$ to $|r\rangle$ and from $|r\rangle$ to $|\!\downarrow\rangle$.}\label{figure1}
\end{center}
\end{figure}

Our experimental setup is depicted in Fig.~\ref{figure1}(a).
Two $^{87}$Rb atoms are held in two optical tweezers separated by 4~$\mu$m. The interatomic axis is aligned with a magnetic field ($B$=9~G), which defines the quantization axis and lifts the degeneracy of the Zeeman sublevels. The tweezers are formed by two laser beams at 810~nm which are sent at a small angle through a microscope objective focusing the beams to a waist of 0.9~$\mu$m. Atoms are captured from an optical molasses and, due to the small trapping volume, either one or no atom is captured in each trap~\cite{Schlosser01}. The same objective collects the fluorescence light of the atoms induced by the molasses beams at 780~nm. The light coming from each trapped atom is directed onto separate avalanche photodiodes (APDs) which allows us to discriminate for each trap whether an atom is present or not.

The relevant levels of $^{87}$Rb are shown in Fig.~\ref{figure1}(b). We have chosen the Rydberg state $|r\rangle=|58d_{3/2},F=3,M=3\rangle$. The interaction energy between two atoms in this state is enhanced by a F\"{o}rster resonance~\cite{Walker08} which leads to a calculated interaction energy $\Delta E/h\approx 50$~MHz for a distance between the atoms of 4~$\mu$m~\cite{Gaetan09}. The qubit ground states  considered for the entanglement are $|\!\downarrow\rangle= |F=1,M=1\rangle$ and $|\!\uparrow\rangle=|F=2,M=2\rangle$ of the $5s_{1/2}$ manifold, separated in frequency by 6.8~GHz. To excite one atom from $|\!\uparrow\rangle$ to $|r\rangle$, we use a two-photon transition with a $\pi$-polarized laser at 795~nm and a $\sigma^+$-polarized laser at 475~nm. The frequency of the 795~nm laser is blue-detuned by $\delta$=600~MHz from the transition from $|\!\uparrow\rangle$ to $(5p_{1/2},F=2)$ in order to reduce spontaneous emission. The measured Rabi frequency of the two-photon transition from $|\!\uparrow\rangle$ to $|r\rangle$ is $\Omega_{\uparrow r}/2\pi\approx 6$~MHz for a single atom. We use the same 475~nm laser for the transition from $|r\rangle$ toÊ $|\!\downarrow\rangle$, but a second 795~nm laser, linearly polarized perpendicular to the quantization axis, with aÊ frequency 6.8~GHz higher to address state $|\!\downarrow\rangle$. The measured Rabi frequency for this second two-photon transition is $\Omega_{r\downarrow}/2\pi\approx 5$~MHz. The two 795~nm lasers are phase-locked to each other using a beat-note technique and  fast electronic correction.
The two lasers are also used to drive Raman rotations between the qubit states $|\!\uparrow\rangle$ and $|\!\downarrow\rangle$. We observe Rabi oscillations between $|\!\uparrow\rangle$ and $|\!\downarrow\rangle$ with an amplitude of 0.95, which includes the fidelity of state initialization and state detection. We set the Rabi frequency of the Raman transition to $\Omega_{\uparrow\downarrow}= 2\pi\times250$~kHz.

We read out the atomic state  by applying a push-out laser beam resonant to the $F$=2 to $F$=3 transition of the D2-line~\cite{Jones07}, which ejects atoms that are in state $|\!\uparrow\rangle$ (or in other $M$-states of the $F=2$ ground level) from the trap. Only atoms that are in $|\!\downarrow\rangle$ (or in other $M$-states of the $F=1$ level) will stay in the trap and will be detected.

\begin{figure}
\begin{center}
\includegraphics[width=14cm]{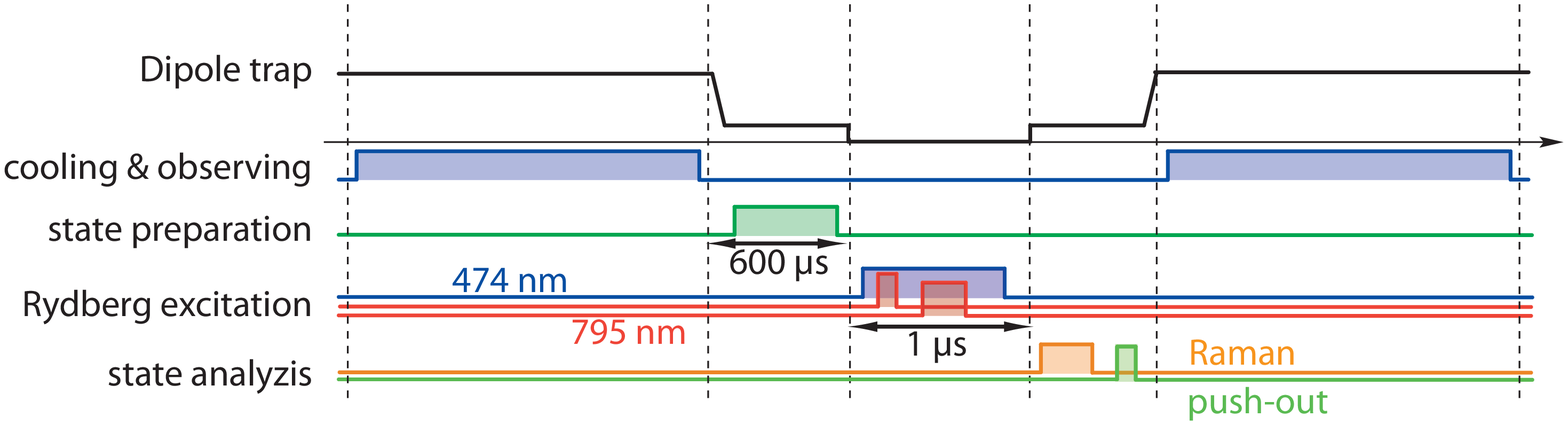}
\caption{Experimental sequence used to entangle two atoms and analyze the entanglement.
The state preparation is done by optical pumping. For clarity the horizontal time axis is not on scale.}\label{expsequence}
\end{center}
\end{figure}

The experimental sequence is shown in figure~\ref{expsequence}.
An experiment starts upon detection of an atom in each trap (trap depth 3.5~mK). After turning off the cooling beams, we ramp adiabatically the trap depth  down to 0.5 mK and optically pump the atoms in $|\!\uparrow\rangle$~\footnote{This reduction of the dipole trap depth decreases the temperature of the atoms  and we have found that it also
leads to a better optical pumping.}.
This is done by a 600~$\mu$s optical pumping phase with a $\sigma^+$-polarized laser coupling the levels $(5s_{1/2}, F=2)$ and $(5p_{3/2}, F=2)$ and a repumping laser from $(5s_{1/2}, F=1)$ to $(5p_{3/2}, F=2)$. Afterwards we switch off the dipole trap while we apply the excitation and mapping pulses towards the Rydberg state and back. The excitation pulse has a duration of $\pi/ (\sqrt{2}\,\Omega_{\uparrow r}) \approx 70$~ns to excite state $|\Psi_{\rm r}\rangle$. The mapping pulse has a duration  $\pi/\Omega_{r\downarrow}\approx 110$~ns.
The trap is then turned on again. In order to analyze the produced two-atom state, we drive global Raman rotations on the two atoms (see below) and the push-out laser is applied. Subsequently, we ramp up the depth of the dipole trap to its initial value
and record for each trap whether the atom is present or not.
We repeat the experiment 100 times for each Raman rotation angle $\theta = \Omega_{\uparrow\downarrow} \tau$ ($\tau$ is the duration of the Raman pulse). We then extract the probabilities  $P_{a}(\theta)$ and $P_{b}(\theta)$ to recapture an atom in trap $a$ or $b$, the joint probabilities $P_{01}(\theta)$ and $P_{10}(\theta)$ to lose atom $a$ and recapture atom $b$ or vice versa, as well as probabilities $P_{11}(\theta)$ and  $P_{00}(\theta)$ to recapture or lose both atoms, respectively, assigning $0$ to a loss and $1$ to a recapture.

Our state-detection scheme, based on the push out technique, identifies  any atom $a$ or $ b$ when it is in  state $|\!\downarrow\rangle$~\footnote{This statement assumes that there is no other Zeeman state of the $(5s_{1/2},F=1)$ manifold populated than $|\!\downarrow\rangle = |5s_{1/2}, F=1, M=1\rangle $. This is indeed the case in our experiment, as explained in section~\ref{sectionlosses}.}.
However, it does not discriminate between atoms in state $|\!\uparrow\rangle$ and atoms that could be lost during the sequence. As a consequence we have to evaluate the amount of these additional losses. We have measured in a separate experiment the probability $p_{\rm recap}$ to recapture a pair of atoms after the excitation and mapping pulses, without applying the push-out laser. We have found  $p_{\rm recap}=0.62(3)$, which shows that
the losses of one or both atoms cannot be neglected. We have incorporated these losses in the analysis of our measurement results, using a model that  is detailed in the next section.

\section{Theoretical model}\label{sectiontheory}

To take into account the loss of atoms, we introduce a set of additional states $\{|x\rangle\}$,
  extending the basis of each atom to
($|\!\uparrow\rangle, |\!\downarrow\rangle, \{|x\rangle\}$)
and we describe the two-atom system by the density matrix $\hat{\rho}$ in this extended
basis. We assume that these additional states $\{|x\rangle\}$, 
corresponding to an atom leaving the qubit basis, 
cannot be distinguished from state $|\!\uparrow\rangle$ by the state detection.
The exact nature of states $\{|x\rangle\}$ will be detailed in section~\ref{sectionlosses}, 
but we already note that in our case they can come either from an atom leaving its trap (physical loss)
or from an atom still trapped but ending up in an unwanted state, 
outside the qubit basis $\{|\!\uparrow\rangle, |\!\downarrow\rangle\}$.
The losses of one or two atoms
are given by the sum of the diagonal elements
$L_{\rm total}=\sum_x (P_{\uparrow x} + P_{\downarrow x} + P_{x \uparrow}+P_{x \downarrow}) +\sum_{x,x'}P_{xx'} $.

We assume that the states  $\{|x\rangle\}$ are   not coupled to $|\!\downarrow\rangle$ or $|\!\uparrow\rangle$ by the Raman lasers, and that they are not coupled between each other.
The Raman rotation for the two atoms can then be described by the operator
$R_{a\otimes b}(\theta,\varphi)= R_a(\theta,\varphi)\otimes R_b(\theta,\varphi)$
where $R_{a/b}(\theta, \varphi)$ is given by the matrix
\begin{equation}
R_{a/b}(\theta, \varphi) = \left( \begin{array}{ccc} \cos{\frac{\theta}{2}} & i
e^{i\varphi} \sin{\frac{\theta}{2}} & 0 \\ 	i e^{-i\varphi}
\sin{\frac{\theta}{2}} Ê& \cos{\frac{\theta}{2}} & 0 \\ 0 & 0 & \hat{1} \\
\end{array} \right)_{|\!\uparrow\rangle,|\!\downarrow\rangle, \{|x\rangle\}}\ , \label{matrice_rotation}
\end{equation}
where $\hat{1}$ stands for the identity matrix, $\theta = \Omega_{\uparrow\downarrow}\tau$ and
 $\varphi$ is the phase difference between the two Raman lasers. 
 The two atoms are exposed to the same laser field and undergo a 
 rotation with the same $\theta$ and $\varphi$. 
 After the rotation the density matrix of the produced state is
$\hat\rho_{\rm rot}(\theta, \varphi)= R_{a\otimes b}(\theta,\varphi)
\hat\rho  R_{a\otimes b}^{\dagger}(\theta,\varphi)$.
The idea behind this approach is  to transform the coherences (off-diagonal matrix element) into populations that can be directly measured.

In our experiment, we do not control the phase $\varphi$ of the Raman lasers with respect to the phase of the atomic states. This comes ultimately from the fact that the atoms are loaded in the dipole traps at random, so that there is no phase relation with respect to the microwave used to generate the Raman transition.
This phase $\varphi$ varies randomly from shot-to-shot over $2\pi$. Our measurement results are therefore averaged over $\varphi$.
When averaging  $\langle\hat\rho_{\rm rot}(\theta, \varphi)\rangle_\varphi$, all
coherences of $\hat\rho_{\rm rot}$ average out, apart from the off-diagonal
element $\rho_{\downarrow\uparrow,\uparrow\downarrow}$ relevant to characterize
state $|\Psi^+\rangle$.
We then  calculate the expressions averaged over $\varphi$ of the probabilities $P_{a/b}(\theta)$ as well as  $P_{11}(\theta)$ and $\Pi(\theta)=P_{11}(\theta)+P_{00}(\theta)-P_{01}(\theta)-P_{10}(\theta)$
as a function of the matrix elements of $\hat{\rho}$.

As our state detection identifies a recapture (1) with the atom being in state $|\!\downarrow\rangle$, we get for the probability to recapture atom $a$ independently of the state of atom $b$:
\begin{eqnarray}\label{Pa}
P_{a}(\theta) &=& P_{\downarrow \downarrow}(\theta)+P_{\downarrow\uparrow}(\theta)+
\sum_{x}P_{\downarrow x}(\theta)\\\nonumber
&=&\frac{1}{2}\left[P_{\uparrow\downarrow}+ P_{\downarrow\uparrow}+ P_{\uparrow\uparrow}+P_{\downarrow\downarrow} +\sum_{x} (P_{\uparrow x}+P_{\downarrow x})\right]\\\nonumber
&+&\frac{1}{2}\left[P_{\downarrow\downarrow}- P_{\uparrow\uparrow}+P_{\downarrow\uparrow}-
P_{\uparrow\downarrow} +\sum_{x} (P_{\downarrow x}-P_{\uparrow x})\right]\cos{\theta}\ .
\end{eqnarray}
In this formula and in the following ones,
$P_{n,m}(\theta)= \langle n,m| \hat{\rho}_{\rm rot}|n,m\rangle$, and $P_{n,m}=P_{n,m}(0)$ with $\{n, m\} \in\{\downarrow,\uparrow,x\}$.
Similarly, the probability to recapture atom $b$ independently of the state of atom $a$ is \begin{eqnarray}\label{Pb}
P_{b}(\theta)&=& P_{\downarrow \downarrow}(\theta)+P_{\uparrow\downarrow}(\theta)+
\sum_{x}P_{x\downarrow }(\theta)\\\nonumber
&=&\frac{1}{2}\left[P_{\uparrow\downarrow}+ P_{\downarrow\uparrow}+ P_{\uparrow\uparrow}+P_{\downarrow\downarrow} + 
\sum_{x} (P_{x\uparrow}+ P_{x\downarrow})\right]\\\nonumber
&+&\frac{1}{2}\left[P_{\downarrow\downarrow}- P_{\uparrow\uparrow}+ P_{\uparrow\downarrow}-
P_{\downarrow\uparrow} +\sum_{x} (P_{x\downarrow }- P_{x\uparrow })\right]\cos{\theta}
\end{eqnarray}
We also introduce the probability $L_{a}$ that atom $a$ lays outside the qubit basis
$\{|\!\uparrow\rangle,|\!\downarrow\rangle\}$, given by $L_{a}=\sum_{x} (P_{x\uparrow}+P_{x\downarrow})+\sum_{x,x'}P_{x,x'}$ and similarly for atom $b$,
$L_{b}=\sum_{x} P_{\uparrow x}+P_{\downarrow x}+\sum_{x,x'}P_{x,x'}$.
From expression~(\ref{Pa}) and~(\ref{Pb}) the probabilities $L_{a}$ and $L_{b}$ are related to the mean value of $P_{a/b}(\theta)$ by the expression
\begin{equation}\label{formuleLa}
\langle P_{a/b}(\theta)\rangle = \frac{1}{2} (1-L_{a/b})\ .
\end{equation}
This expression is intuitive:  the mean value of the probability for an atom to be recaptured, i.e. the atom is in state $|\!\downarrow\rangle$, is 1/2 when there is no additional loss during the entangling sequence. When we take into account the probability to lose the atom, we simply multiply the probability in the absence of additional loss, 1/2, with the  probability to stay in the qubit basis $1-L_{a/b}$.

The calculation gives the joint probability to recapture both atoms at the end
of the Raman rotation:
\begin{eqnarray}\label{P11}
P_{11}(\theta) &=& P_{\downarrow \downarrow}(\theta)\\\nonumber
&=&\frac{1}{8}\left[P_{\uparrow\downarrow}+
P_{\downarrow\uparrow}+
2\Re(\rho_{\downarrow\uparrow,\uparrow\downarrow})+
3(P_{\uparrow\uparrow}+P_{\downarrow\downarrow})\right]\\\nonumber
&+&\frac{1}{2}(P_{\downarrow\downarrow}-P_{\uparrow\uparrow})\cos{\theta}\\\nonumber
&+&\frac{1}{8}\left[P_{\downarrow\downarrow}+
P_{\uparrow\uparrow}-P_{\uparrow\downarrow}-P_{\downarrow\uparrow}-2\Re(\rho_{\downarrow\uparrow,\uparrow\downarrow})\right]\cos{2\theta}\ .
\end{eqnarray}
Here, ${\Re }$ denotes the real part. This expression exhibits terms oscillating at frequencies $\Omega_{\uparrow\downarrow}$
and $2\Omega_{\uparrow\downarrow}$. The term at $\Omega_{\uparrow\downarrow}$
reflects the imbalance between the states $|\!\uparrow,\uparrow\rangle$ and $|\!\downarrow,\downarrow\rangle$. We note also that this expression of $P_{11}$ does not involve any loss terms, as it characterizes situations where both atoms are present at the end of the sequence. That is why we focus on this quantity for extracting the amount of entanglement between the two atoms.

Finally, we calculate the signal  $\Pi (\theta )$, which is is equal to the parity~\cite{Turchette98} when there are no losses from the qubit basis. We find the expression:
\begin{eqnarray}\label{Parite}
\Pi({\theta} )&=& \frac{1}{2}\left[P_{\downarrow\downarrow}+P_{\uparrow\uparrow}
-P_{\uparrow\downarrow}-P_{\downarrow\uparrow}
+2\Re(\rho_{\downarrow\uparrow,\uparrow\downarrow})+2\sum_{x,x'}P_{xx'}\right]\\\nonumber
&+&\sum_{x} (P_{x\uparrow}+P_{\uparrow x}-P_{x\downarrow}-P_{\downarrow x})\cos{\theta}\\\nonumber
&+&\frac{1}{2}\left[P_{\downarrow\downarrow}+P_{\uparrow\uparrow}
-P_{\uparrow\downarrow}-P_{\downarrow\uparrow}-
2\Re(\rho_{\downarrow\uparrow,\uparrow\downarrow})\right]\cos{2\theta}
\end{eqnarray}
This formula also presents oscillations at two frequencies, the one at $\Omega_{\uparrow\downarrow}$ being related this time to events where only one of the two atoms are present.

As a final remark on this model we  point out that a global rotation with no control over
the phase $\varphi$ would not be suitable to analyze the Bell states
$|\Psi^{-}\rangle=( |\!\uparrow,\downarrow\rangle - |\!\downarrow,\uparrow\rangle)/\sqrt{2}$
and $|\Phi^{\pm}\rangle=( |\!\uparrow,\uparrow\rangle \pm |\!\downarrow,\downarrow\rangle)/\sqrt{2}$.
As an example, the antisymetric state $|\Psi^{-}\rangle$ does not change under the rotation~\cite{Turchette98}, whatever the
phase $\varphi$. For the states $|\Phi^{\pm}\rangle$, the coherence
$\rho_{\downarrow\downarrow,\uparrow\uparrow}$ acquires under the rotation a phase
factor $e^{-i2\varphi}$. On a single realization of the experiment, the phase
$\varphi$ is fixed but the average over many realizations  cancels out. The robustness of  $|\Psi^+\rangle$
under fluctuations of  $\varphi$ is reminiscent of the fact that this state lies in a decoherence free subspace~\cite{Haffner05}.

In the remaining part of the paper, we will use this model to extract from a single set of data the probability to lose one and two atoms, as well as the amount of entanglement.

\section{Analysis of the losses}\label{sectionlosses}

\begin{figure}
\begin{center}
\includegraphics[width=10cm]{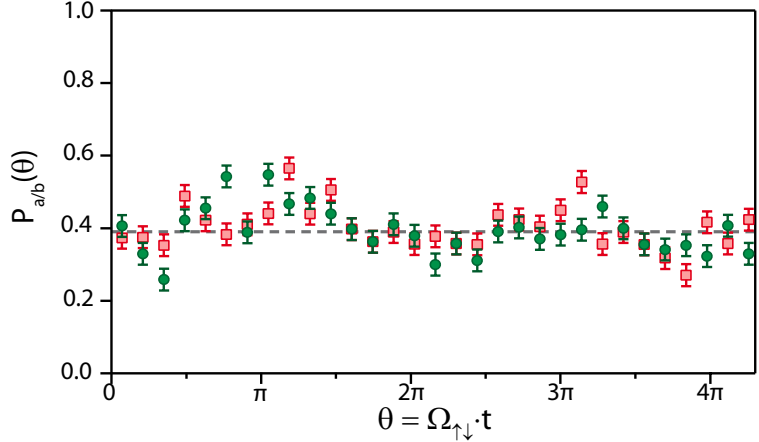}
\caption{Measured probabilities $P_{a}(\theta)$ (red squares)
and $P_{b}(\theta)$ (green dots)
to recapture each atom at the
end of the entanglement procedure, followed by a Raman pulse on
both atoms for different
pulse durations. The dotted indicates the mean value of 
$P_{a/b}(\theta)$.}\label{figPab}
\end{center}
\end{figure}

Figure~\ref{figPab} shows the recapture probabilities $P_{a/b}(\theta)$ for each atom for different values of the Raman rotation angle. From equation~(\ref{formuleLa}) and the mean value of $P_{a/b}(\theta)$ deduced from the data we find $L_a=L_b = 0.22(1)$, confirming  that  the loss probability is the same for both atoms.
Assuming independent losses for atoms $a$ and $b$ we find the probability to lose at least one of the two atoms $L_{\rm total}=L_a(1-L_{b}) + L_b(1-L_{a}) + L_a L_b = 0.39(2)$.
The recapture probability of a pair of atoms in the qubit basis $\{|\!\uparrow\rangle,|\!\downarrow\rangle\}$ is then $L_{\rm total}= 1-{\rm tr}\hat{\rho}$, restricting the trace to pairs of atoms still present at the end of the entangling sequence in the states $|\!\uparrow\rangle$ and $|\!\downarrow\rangle$.

The loss channels can be separated in three classes.
In the first category, independent of the Rydberg excitation and Raman rotation, we measured losses during the time the trap is switched off ($\sim 3\%$) as well as
errors in the detection of the presence of the atom ($\sim 3\%$). For this first category, the loss channels $\{|x\rangle\}$ correspond to an atom in any internal state but which is lost from the tweezers.

In the second category, the losses are also physical and
occur during the entangling and mapping pulses. These losses 
correspond to situations where one or two atoms have left the dipole traps, 
and are therefore absent when the Raman rotation and the measurement take place.
These losses are independent of the state detection and are mostly related to the fact that an 
atom left in the Rydberg state is lost, since  it is not trapped in the dipole trap. 
Using a model based  on Bloch equations including the 5 relevant states 
($|\!\uparrow\rangle$, $|\!\downarrow\rangle$, $|5s_{1/2}, F=2,M=1\rangle$, $|5p_{1/2}, F=2,M=2\rangle$ 
and $|r\rangle$), we identify the following scenarios.
Firstly, spontaneous emission from the $5p_{1/2 }$ state populates the state  $|\!\downarrow\rangle$ from which $\sim 7~\%$ of the atoms get excited to the Rydberg state by the mapping pulse.
Secondly, intensity fluctuations ($5~\%$) and frequency fluctuations ($3$~MHz) of
the excitation lasers reduce the efficiency of the mapping pulse so that $\sim 7~\%$ of the atoms will not be transferred back from the Rydberg state to $|\!\downarrow\rangle$.
For this second class, the loss channel $|x\rangle$ is any Rydberg states $|r\rangle$
which can be coupled by the two-photon transition including the one resulting,
e.g. from an imperfect polarization of the lasers.

The third class of losses corresponds to atoms that are still present at the end of the entangling
and mapping sequence, but which are in states different from $|\!\uparrow\rangle$
and $|\!\downarrow\rangle$, that is outside the qubit basis when the state
measurement is performed. Because of the selection rules, the main possibility in our case is  the spontaneous emission from the $5p_{1/2}$ to state $|x\rangle=|5s_{1/2}, F=2,\,M=1\rangle$ during the entangling and mapping pulses which is calculated to be only
$\sim 2 \%$, due to a small branching ratio.
This third contribution is therefore small in our case. 
By adding the contributions of the three categories of losses we find 
a loss probability for each atom of 0.22, in agreement with the measured values 
of $L_{a/b}$.

Finally, we compare the probability $1-L_{\rm total}=0.61(2)$ for  a pair of atoms to be 
in states $|\!\uparrow\rangle$ or $|\!\downarrow\rangle$ with 
the probability $p_{\rm recap}=0.62(3)$ to recapture both atoms, 
irrespective of their internal states. 
Both values are almost identical, confirming that the dominant mechanism is a 
physical loss of the atoms before the state measurement.

\section{Partial state reconstruction and fidelity}\label{sectionfidelity}

\begin{figure}
\begin{center}
\includegraphics[width=8.6cm]{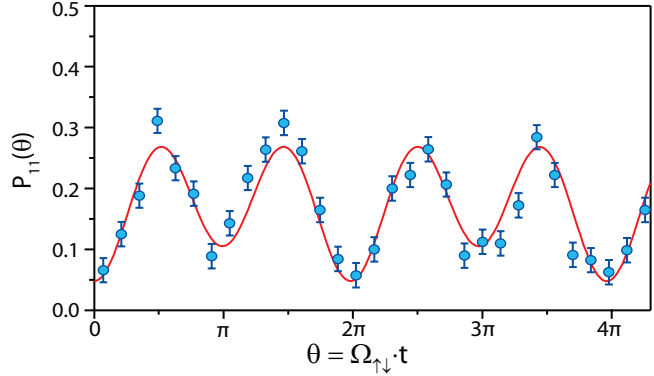}
\caption{Measured probability $P_{11}(\theta)$ to recapture the two atoms at the
end of the entanglement procedure, followed by a Raman pulse on both atoms for different
pulse durations. The data are fitted by a function of the form $y_{0}+A\cos \Omega_{\uparrow\downarrow} t + B\cos 2 \Omega_{\uparrow\downarrow} t$, according to the discussion in the text. The error bars on the data are statistical.
The fit gives $y_{0}=0.17(2)$, $A=-0.03(1)$ and $B=-0.096$.}\label{figP11}
\end{center}
\end{figure}

In order to analyse the two-atom state 
we focus on the the joint recapture probability for atom pairs $P_{11}(\theta)$ shown in figure~\ref{figP11}, since it incorporates no loss terms, as shown in equation~(\ref{P11}).
For the maximally entangled state $|\Psi^{+}\rangle$, $P_{11}(\theta)$
should 
oscillate between 0 and $1/2$ at a frequency $2\Omega_{\uparrow\downarrow}$, 
while here the data show oscillations at two frequencies $\Omega_{\uparrow\downarrow}$  and
$2\Omega_{\uparrow\downarrow}$, with a reduced amplitude.
From the measurement of $P_{11}(\theta)$ and the expression~(\ref{P11}), we extract  $P_{\downarrow\downarrow} = P_{11}(0)$ and $P_{\uparrow\uparrow} = P_{11}(\pi)$. Combining the value of the total losses $L_{\rm total}$ and the normalization condition  $P_{\uparrow\downarrow} +P_{\downarrow\uparrow}+P_{\uparrow\uparrow} +P_{\downarrow\downarrow}+ L_{\rm total}=1$,
we get $P_{\uparrow\downarrow} +P_{\downarrow\uparrow}$.
The mean value  $\langle P_{11}(\theta)\rangle = [P_{\downarrow\uparrow} + P_{\uparrow\downarrow} +  3 P_{\downarrow\downarrow} + 3P_{\uparrow\uparrow} +2{\Re} (\rho_{\downarrow\uparrow,\uparrow\downarrow})] / 8$ yields $\Re(\rho_{\downarrow\uparrow,\uparrow\downarrow})$.
Table~\ref{summary} summarizes the complete information about the density
matrix $\hat\rho$ one can extract from global Raman rotations without control of $\varphi$.

\begin{table}
\begin{center}
\begin{tabular}{c c c}
\hline
\hline
Matrix elements & &Experimental values\\
\hline
$\rho_{\downarrow\downarrow,\downarrow\downarrow}=P_{\downarrow\downarrow}$ & &$0.06\pm 0.02$\\
$\rho_{\uparrow\uparrow,\uparrow\uparrow}=P_{\uparrow\uparrow}$ & &$0.09\pm 0.02$\\
$\rho_{\downarrow\uparrow,\downarrow\uparrow}+\rho_{\uparrow\downarrow,\uparrow\downarrow} =P_{\downarrow\uparrow}+P_{\uparrow\downarrow}$& &$0.46\pm 0.03$\\
${\Re}(\rho_{\downarrow\uparrow,\uparrow\downarrow}) $& &$0.23\pm 0.04$\\
\hline
\hline
\end{tabular}
\caption{Measured values of the density matrix elements characterizing the state prepared in the experiment extracted from $P_{11}(\theta)$. The error bars are statistical. Note that the restriction to the states $|\!\uparrow\rangle$ and $|\!\downarrow\rangle$ leads to ${\rm tr}(\hat\rho) = 0.61$
because of the loss $L_{\rm total} = 0.39(2)$ from the qubit basis.}\label{summary}
\end{center}
\end{table}

\begin{figure}
\begin{center}
\includegraphics[width=8.6cm]{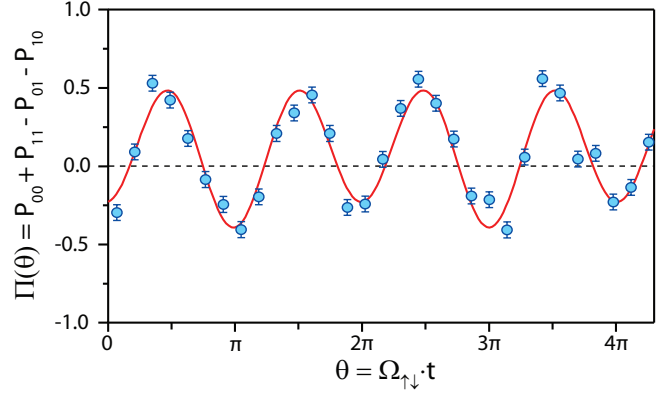}
\caption{Measured signal $\Pi(\theta)$ for different durations of the analysing Raman pulse. The data are fitted by a function of the form $y_{0}+ A\cos \Omega_{\uparrow\downarrow} t + B\cos 2 \Omega_{\uparrow\downarrow} t$ as discussed in the text. The error bars on the data are statistical.
The fit gives $y_{0}=0.08(1)$, $A=-0.07(1)$ and $B=-0.39(1)$.}\label{figparity}
\end{center}
\end{figure}

As a cross-check of our data analysis, we look at the signal $\Pi (\theta )$ which is shown in figure~\ref{figparity}.  For the maximally entangled state $|\Psi^{+}\rangle$,
the parity should 
oscillate between $-1$ and $+1$ with a frequency  of $2\Omega_{\uparrow \downarrow}$, 
while here the observed $\Pi (\theta )$ oscillates at two frequencies,
$\Omega_{\uparrow\downarrow}$ and $2\Omega_{\uparrow\downarrow}$ with reduced amplitude.
From equation~(\ref{Parite}) we calculate $\Pi (\pi/2) = 2{\Re} (\rho_{\downarrow\uparrow,\uparrow\downarrow}) + \sum_{x,x'}P_{xx'}$.
Under the assumption that losses are independent for atoms $a$ and $b$, as mentioned in section~\ref{sectionlosses}, 
$L_{\rm total}=L_a + L_b - L_a L_b$. 
Combining this formula with the expressions of $L_{\rm total}$, $L_{a}$ and $L_{b}$ given in section~\ref{sectiontheory}, we get $\sum_{x,x'}P_{xx'}=L_{a} L_{b}$.
We then deduce the coherence ${\Re}(\rho_{\downarrow\uparrow,\uparrow\downarrow}) = 0.22(4)$, which is in good agreement with the value deduced from the analysis of $P_{11}(\theta)$ described above.

Our analysis allows us to calculate the fidelity of the 
entangling operation. This  fidelity $F$ is defined by $F=\langle\Psi^+ |\hat\rho| \Psi^+\rangle = 
(P_{\downarrow\uparrow} + P_{\uparrow\downarrow})/2 + 
{\Re} (\rho_{\downarrow\uparrow,\uparrow\downarrow})$ 
with respect to the expected $|\Psi^+\rangle$ Bell 
state~\cite{Sackett00}.
From the values in table~\ref{summary},
we get $F=0.46(4)$.  
This fidelity $F$ is defined with respect to the initial number 
of atom pairs and includes events for which one or two atoms have been 
lost physically during the entangling sequence.
That means $F$ characterizes the whole entangling 
operation which is mainly limited by atom losses. 
As $F<0.5$, this value does not prove entanglement between the atoms.

The quantum nature of the correlations between the two atoms is revealed 
if we calculate the fidelity $F_{\rm pairs} = F/p_{\rm recap}$ 
which characterizes the pairs of atoms effectively present at the end of the entangling sequence
before state detection. 
From $p_{\rm recap}=0.62(3)$, we calculate $F_{\rm pairs} = 0.74(7)$. 
This approach to take into account atom losses, is very similar to the one used 
in Bell inequality tests with photons based on {\it one-way polarizers}~\cite{Freedman72, Aspect81}. 
In these experiments, 
the absence of a photon detection after the polarizer can be due to a 
photon with orthogonal polarization, or a photon that has been 
lost before reaching the polarizer.
Therefore, the total number of
detected photon pairs is first measured by removing the polarizers, then the  measurement 
of the polarization correlation is performed and the results are renormalized by the 
total number of photon pairs.

Our analysis gives also access to the fidelity $F_{\uparrow\downarrow} = F/{\rm tr}\hat{\rho}$ which characterizes the entanglement of atom pairs which are still in the qubit basis $\{|\!\uparrow\rangle,|\!\downarrow\rangle\}$. We find $F_{\uparrow\downarrow}=0.75(7)$, which is very close to $F_{\rm pairs}$ since the main mechanism for atom losses is the physical loss of one or two atoms from their traps. The fact that  $F_{\rm pairs}>0.5$ and $F_{\uparrow\downarrow} > 0.5$
proves that the two atoms are entangled.
We can identify two effects lowering the fidelity  with respect to the ideal case. Firstly, an imperfect Rydberg blockade leads to the excitation of both atoms (probability $\sim 10\%$~\cite{Gaetan09}) and their subsequent mapping to the state $|\!\downarrow,\!\downarrow\rangle$, resulting in a non-zero component of $P_{\downarrow\downarrow}$. Secondly, the excess value of $P_{\uparrow\uparrow}$ is explained by spontaneous emission  from the state $5p_{1/2}$ as well as imperfect Rydberg excitation from the two atom state $|\!\uparrow,\!\uparrow\rangle$. We note that in the present status of the experiment, the influence of the residual motion of the atoms
in their traps is negligible on the fidelity (for more details, see~\cite{Wilk09}).

\section{Conclusion}
In conclusion, we have used global Raman rotations to analyze the entanglement of two atoms which is created using
the Rydberg blockade. Our analysis is based on a model taking into account losses of atoms. We have found that the 62\%  pairs of atoms  remaining at the end of the sequence are in a state with a  fidelity 0.74(7) with respect to the expected $|\Psi^+\rangle$, showing the non-classical origin of the correlations. Future work will be devoted to the measurement of the coherence time of the entangled state, as well as to the improvement of the fidelity and the state detection scheme.

\begin{acknowledgments}
We thank M. Barbieri, M. M\"uller, R. Blatt, D. Kielpinski and P. Maunz for
discussions. We acknowledge support from the European Union through the Integrated Project SCALA, IARPA and the Institut Francilien de Recherche sur les Atomes Froids (IFRAF). A. Ga\"{e}tan and C. Evellin are
supported by a DGA fellowship. T. Wilk is supported by IFRAF.
\end{acknowledgments}

\end{document}